Predicting COVID-19 distribution in Mexico through a discrete and time-dependent Markov chain and an SIR-like model.

Alfonso Vivanco-Lira[1,2].

[1]Undergraduate medical student.

Medical Sciences Department.

University of Guanajuato.

Leon, Guanajuato, Mexico.

[2]Undergraduate mathematics student.

Exact Sciences and Engineering Division.

Open and Distance Learning University of Mexico.

Mexico City, Mexico.

Email: poncho9715@gmail.com

## *Abstract.*

COVID-19 is an emergent viral infection which rose in December 2019 in a city in the Chinese province of Hubei, Wuhan; the viral aetiology of this infection is now known as COVID-19 virus, which belongs to the *Betacoronavirus* genus. This virus produces the syndrome of acute respiratory stress that h as been witnessed in other coronaviruses, such as that MERS-CoV in Middle East countries or SARS-CoV which was seen in 2002 and 2003 in China. This virus mediates its entry through its spike (S) proteins interacting with ACE2 receptors in lung epithelial cells, and may promote an inflammatory response by means of inflammasome NLRP3 activation and unfolded protein response (these are possibly consequence of the envelope E protein of COVID-19 virus). Efforts have been made worldwide to prevent further spread of the disease, but in March 2020 the WHO declared it a pandemic emergency and Mexico started to report its first cases. In this paper we attempt to summarize the biological features of the virus and the possible pathophysiological mechanisms of its disease, as well as a stochastic model characterizing the probability distribution of cases in Mexico by states and the estimated number of cases in Mexico through a differential equation model (modified SIR model), thus will we be able to characterize the disease and its course in Mexico in order to display more preparedness and promote more logical actions by both the policy makers as well as the general population.

Keywords.

COVID19, SARS, spike, envelope, Markov, SIR.

## *Introduction.*

Coronaviruses are known for causing respiratory (if not gastrointestinal or neurological) diseases in both humans and animals, however before 2002 and 2003 outbreaks of severe acute respiratory syndrome (SARS) in China, this set of viruses was thought to cause mainly affection in animals (Cui, Li, & Shi, 2019), for it has been calculated that annually $40 x 10^9$ *Gallus gallus domesticus* specimens in the world suffer from infectious bronchitis coronavirus and that 90% of the infected swine by porcine transmissible gastroenteritis coronavirus die (Cavanagh, 2005). After the 2002-2003 outbreak of SARS in China, another epidemiologically relevant coronavirus infection rose in Middle East countries represented by the Middle East respiratory syndrome coronavirus (MERS-CoV); finally, by the end of 2019, an increase in pneumonias of unknow origin was revealed in Wuhan, a city in the province of Hubei in China, which were later revealed to be etiologically related by a novel coronavirus, say 2019-nCoV.

## *Biology and pathophysiology of coronaviruses.*

Coronaviruses are members of the *Nidovirales* order according to the International Committee on Taxonomy of Viruses (de Groot, et al., 2018) belonging to the *Cornidovirineae* suborder, being themselves a family of viruses, the *Coronaviridae* in which we may encounter two subfamilies: *Letovirinae* and *Orthocoronavirinae*; it is within the latter subfamily of viruses that we find the genera: *Alphacoronavirus, Betacoronavirus, Deltacoronavirus* and *Gammacoronavirus*. SARS-2003-related coronavirus is found in the subgenus *Sarbecovirus* of *Betacoronaviruses*; while MERS-CoV is found in the subgenus *Merbecovirus* of *Betacoronaviruses*. An alternative classification was proposed in which coronaviruses are grouped in tetramers: Group 1 (TGEV, FCoV, CECoV, PEDV, HCoV-229E), Group 2 (HEV, BCoV, CRCoV, MHV, RtCoV, PuCoV, HCoV-OC43), Group 3 (IBV, TCoV, PhCoV) and Group 4 (SARS-CoV) (Cavanagh, 2005). Coronaviruses are enveloped positive-sense RNA viruses and display the largest genome sizes amongst all RNA viruses (from 26.2 to 31.7 kb (McBride, van Zyl, & Fielding, 2014). The virions have diameters that range from 80 to 120 nm, however smallest as 50 nm and largest as 200 nm diameters have been noted. From the envelope some proteins project from 15 to 20 nm, these are spike (S) proteins and all coronaviruses have genes to code these proteins (S gene) which internalize in the virion and are mechanically bound by means of a "foot" like extension of 10 nm long. Along with these spike proteins, there exist shorter proteins which protrude the virion surface from 5 to 10 nm, these being the hemagglutinin-esterase (HE) proteins (Masters, 2006). The nucleocapsid protein (CoVN) shows conserved regions: N terminal domain (NTD), C-terminal domain (CTD) and an intrinsically disordered central region (RNA-binding domain); however the name of the domains, all three of them have been shown to interact with viral RNA. The NTD has shown a grade of divergence amongst coronaviruses both in amino acid content and length, this domain associated with the 3' end of the viral RNA genome by means of Gaussian electrostatic interactions; the NTD shows convergence in some structural aspects: secondary structures, such as the beta sheet flanked by alpha helices, a basic RNA binding groove, enrichment in aromatic and basic residues. The NTD resembles "hand"-like structure: an acidic wrist, a hydrophobic palm and basic fingers. The CTD domain is hydrophobic and promotes homodimerization of the nucleoprotein (sometimes, not only homodimerization but also oligomerization). This protein promotes the formation of the viral capsid protecting the virus from degradation by extracellular agents; the self-association of the N protein has been seen as a target in coronavirus infection treatment (McBride, van Zyl, & Fielding, 2014). Within the coronavirus genome we may find the polyproteins, which show replicase activity; this protein uses the genome as the template for the synthesis producing mRNAs, these are then translated into structural proteins and others; the structural proteins transit then hatch to the endoplasmic reticulum (ER) transiting then towards the ER-Golgi intermediate compartment (ERGIC). The nucleocapsids are formed when the progeny genomes interact with the N protein; these newly formed nucleocapsids bud into the ERGIC, forming then virions ready to exit the cell (Masters, 2006). Coronaviruses' spike proteins have been shown to interact with cellular receptors for their internalization into the cell, such as the human aminopeptidase N protein (ANPEP, membrane alanyl aminopeptidase) and the angiotensin-converting enzyme 2 (ACE2) (Masters, 2006).

*CoV proteins.*

*Spike (S) protein.*

This protein contains three segments: a large ectodomain (which consists of a subunit S1 with receptor-binding activity and an S2 subunit with membrane-fusion activity), a transmembrane anchor and an intracellular tail. Trimerization of the protein is in order when seen at the microscope having S1 subunits in one extracellular extreme and S2 subunits in the other extreme. Various domains of the spike protein bind to different proteins/receptors in the host: S1-N-terminal domain binds to sugars (in *Alphacoronavirus*), to CEACAM1 (in MHV *Betacoronavirus*), and sugars in BCoV and IBV (a *Gammacoronavirus*). Whereas the S1-C-terminal domain binds to ACE2 (in *Alphacoronaviruses* and SARS-CoV), aminopeptidase N (TGEV-PEDV, PRCV), and DPP4 (in MERS-CoV). The S1 subunits display divergence in their sequences across genera, however convergence is seen in those of the same genus. The S1 C-terminal domain subunit contains itself two subdomains (as it has been seen in that S protein corresponding to SARS-CoV) these subdomains are: a core structure and a receptor-binding motif, this latter subdomain binds to ACE2 by means of a concave outer surface, this concavity is formed by a beta-sheet with two ridges (loops). Two virus-binding residues in ACE2 have been shown to highly contribute to the binding of the virus by promoting a regional hydrophobic environment: Lys31 and Lys353; this specificity can tell us about how a couple of mutations in both ACE2 or the S1 subunit in the spike protein may promote either resistance (or increased susceptibility) or species tropism in regards to the infection. This S1-C-terminal subunit in MERS-CoV binds to DPP4, this subunit contains as well, in itself, two subdomains: a core structure and a receptor binding motif, however this motif is not concave but flat, and DPP4 forms a homodimer, each monomer with a hydrolase domain and a beta-propeller domain (Li, 2016), said structure (beta propellers) have been seen in various species, and several classes are known, depending on the symmetry axis, for instance, the neuraminidase protein of the influenza virus has six four-stranded beta sheets; this beta propeller structure is a highly symmetrical structure with four to eight-fold repeats of a four-stranded antiparallel beta-sheet motif, this beta sheets are arranged around a central tunnel, and a hydrophobic character of the interactions provides the stability (Pons, Gómez, Chinea, & Valencia, 2003), in the case of DPP4, the viral binding motifs are seen in the other surface of the beta propeller domain (Li, 2016). It has been noted that the COVID19 S2 subunit displays a high degree of conservation, contrary to the S1 subunit with 40% of identity with other SARS-CoVs (Cascella, Rajnik, Dulebohn, & Di Napoli, 2020). The spike protein has been considered a member of the class I viral membrane fusion proteins, including those from: influenza virus, human immunodeficiency virus and Ebola virus; in which we may see: hemagglutinin glycoprotein corresponding to the influenza virus, Env protein in HIV (trimer of gp120 and gp41 heterodimers (Wilen, Tilton, & Doms, 2012)), and GP (glycoprotein) in Ebola virus (which is a triplet of heterodimers, composed of a receptor binding subunit GP1, and a fusion subunit GP2) (Salata, et al., 2019). Analogies can be made from the fusion process corresponding to influenza virus and hemagglutinin, for which there exist both a prefusion and a postfusion state, in the latter one there exists a dramatic conformational change, fusion peptides are exposed and insert into the target membrane, there is an energy barrier for such a conformational change which may be overcome by means of receptor binding (as seen in HIV fusion), low pH (influenza virus) or both (avian leucosis virus); this energy barrier in coronavirus is thought to be worked out by means of proteolysis of the spike proteins: proprotein convertases, during virus packaging, extracellular proteases, cell surface proteases, lysosomal proteases; low pH and receptor binding (as seen in other class I fusion proteins) may as well promote the reaction and overcome the energy barrier (Li, 2016).

*Hemagglutinin-esterase (HE) protein.*

This protein is exclusive to those corresponding to the group 2a of coronaviruses (Zeng, Langereis, van Vliet, Huizinga, & de Groot, 2008) (mouse hepatitis virus, rat coronavirus, human respiratory coronaviruses OC43 (HCoV-OC43), HKU1, canine respiratory CoV, porcine hemagglutinating encephalomyelitis virus, equine coronavirus, bovine coronavirus and wild-ruminant coronaviruses (Hasoksuz, Vlasova, & Saif, 2008)), toroviruses and orthomyxoviruses; this would indicate that RNA recombination and genetic transfer may have occurred in the past. In coronaviruses, HE does not possess fusion activity and is merely accessory to the spike protein, nevertheless, they do bind to sialic acid and have shown to have receptor-destroying activity. The CoV HE has three domains: a receptor binding domain, an acetylesterase domain, and a membrane-proximal domain (the thrice show equivalence with those of HEF influenza protein) (Zeng, Langereis, van Vliet, Huizinga, & de Groot, 2008).

*Envelope (E) protein.*

It is an integral membrane protein of 76 to 109 amino acids with a molecular weight ranging from 8.4 to 12 kDa; this protein has a short hydrophylic N-terminus followed by a large hydrophobic transmembrane domain, ending with a hydrophilic C-terminus; the transmembrane domain contains $\geq 1$ alpha-helix which may oligomerize and forms an ion-conductive pore. It has been also noted that mutations to the PDZ-binding motif (PBM) are endured and may show value in vaccine development. It may have a function in virion budding as it interacts with the Golgi apparatus. It has showed interesting protein interactions with hosts, such as the one noted with BCLXL (BCL2 like 1), PALS1 (MPP5), Na/K ATPase alpha1 subunit, stomatin. Because of its interactions with the endoplasmic reticulum, this has been linked to enhancement of the unfolded protein response (UPR) during infection; and due to the fact that the envelope protein promotes the formation of viroporins, transport of calcium by the E protein promotes inflammosome activation (Schoeman & Fielding, 2019).

*Host proteins.*

*ACE2.*

This gene encodes the angiotensin I converting enzyme 2, which is a member of the angiotensin-converting enzyme family of dipeptidyl carboxypeptidases showing homology to human angiotensin I converting enzyme. This enzyme catalyzes the cleavage of angiotensin I into angiotensin (1-9) and angiotensin II into angiotensin (1-7); it is as well a receptor for the novel COVID19-virus. The gene encoding for this protein is found in Xp22.2 and has 21 exons. This protein is expressed in the following tissues: colon, duodenum, fat, gall bladder, heart, kidney, liver, testis, thyroid and lung amongst others (NIH, 2020), the tissue expression of ACE2 is augmented in patients with type 1 or type 2 diabetes, hypertension, and those who are under treatment with thiazolidinediones and ibuprofen (Fang, Karakiulakis, & Roth, 2020) however those who have not yet received treatment show lower expression levels, this mechanism is thought to be mediated by angiotensin II via ERK/p38 MAP or ERK1/2 and JNK, furthermore, should we block AT1R, ACE2 levels rise, this rise can also be obtained by administering low doses of aldosterone, and decreases are seen when treated with endothelin 1 (Clarke & Turner, 2011); as well interethnic differences in expression patterns have been noted which may predispose to different susceptibilities (Cao, et al., 2020). One of the catalysed products of ACE2 is angiotensin (1-7) which serves to activate MAS1 proto-oncogene, which is a G protein-coupled receptor, its actions include: vasodilation, vascular protection, anti-fibrotic, anti-proliferative and reduction in pro-inflammatory cytokines (Clarke & Turner, 2011). Thus, we may consider ACE2 production pathways-blockage in order to diminish the probability of COVID19-virus to be received and internalized by the cell, by means of JNK (MAPK8), MAPK14 (p38), endothelin-1, AT1R agonists or aldosterone receptors antagonists (spironolactone, eplerenone). ACE2 has been shown to interact itself with two small molecules: moexipril and lisinopril, which may induce conformational changes in ACE2 necessary to diminish the hydrophobicity of the region responsible of the interaction with the spike (S) proteins of coronaviruses (TyersLab, 2020), other chemicals have been noted: 2-amino-N,N-diethyl-1,3-benzothiazole-6-carboxamide (also displaying anti-HIV reverse transcriptase activity and antibacterial activity against *Mycobacterium tuberculosis*) (PubChem, 2020), losartan, D-glucose, streptozotocin, captopril, diminazene aceturate (PubChem, 2020).

*BCLXL (BCL2 like 1).*

It belongs to the BCL2 protein family, which may homo- or heterodimerize and may either promote or suppress apoptosis regulators; this protein may locate at the outer mitochondrial membrane and may regulate the opining of the outer mitochondrial membrane channel thus being able to change mitochondrial membrane potential, reactive oxygen species production and release of cytochrome c, inducers of apoptosis. It shows two isoforms, the longer is antiapoptotic and the shorter is a proapoptotic (NIH, 2020), and it has been seen that BCLXL interacts with the envelope E protein of coronaviruses, enabling then the apoptosis of T cells, compatible with the clinical features of lymphopoenia seen in coronaviruses infections (Yang, et al., 2005).

*Pathophysiology.*

Until now, we have unravelled two clues which may lead to the development of a solid pathophysiological theory: promotion of inflammasome formation and increase in the unfolded protein response, as well as changes in the apoptotic range of certain cells (such as T cells), in regards to the promotion of inflammasome formation, should there be an activation of the NLRP1 inflammasome by means of NLRP1's cleavage, inducing NLPR1B, the latter would induce ASC protein or PYCARD protein (PYD and CARD domain containing protein), promoting the activation of caspase 1 and the consequent activation of interleukin 1 B and interleukin 18 or a pyroptotic death (Broz & Dixit, 2016). However it has been seen that the E protein from SARS viruses induce NLRP3 inflammasome response (Chen, Moriyama, Chang, & Ichinohe, 2019) perhaps by means of the interaction there is sustained with BCLXL and pore formation in mitochondria, which may induce a potassium efflux (reducing the amounts of intracellular potassium) leading to NLRP3 activation, this aided by means of NEK7 which has been shown as a driving factor in cellular division, where NEK7 increased levels diminish mitotic rate and inflammasome activation and caspase 1 activation, also leading then to IL1B and IL18 concentration increases (Broz & Dixit, 2016), the NLRP3 inflammasome activation has been related not only to recruitment of caspase 1 but also of caspase 4 and 5 (PubChem, 2020), moreover, NLRP3 is an upstream activator of NFKB (NIH, 2020); because of these various interactions that NLRP3 shows and the possible clinical signs and symptoms we may encounter due to this activation, it could be advisable in certain cases to induce NLRP3 inactivation in order to diminish respiratory distress by means of small molecules (PubChem, 2020). In regards to IL1B signalling, it has been seen that it interacts with its receptor IL1R, activating IRAK4 (interleukin 1 receptor associated kinase), activating TRAF6 (TNF receptor associated factor 6) and some other downstream proteins (TAB2, TAK1, MKK3, p38), finally promoting the activation of IL6, IL8, as well as COX2, NFKB (Weber, Wasiliew, & Kracht, 2010); thus we may see as well that the restrictions in oxygenation seen in SARS may be very well induced by means of an overt inflammation by COVID19-virus. In turn, the unfolded protein response pathway that is seen activated by means of the envelope protein promotes the activation of NFKB and MAPK8, XBP1u (unspliced X box-binding protein 1) which promotes UPR target genes transcription, EIF2A (this factor inhibits translation and activates ATF4, augmenting UPR target genes transcription), AT6F (also promotes UPR target genes); the phenotype of the UPR target genes transcription is: autophagy, apoptosis, co-translational degradation, protein secretion, lipid synthesis (Hetz, 2012).

## Model.

This model consists of two portions: a discrete time discrete Markov chain which shall predict the probability distribution of COVID19 in time in the Mexican population, for said chain we require the knowledge or the estimation of the infected population in given time, information that shall be known by the SIR (susceptible, infected, recovered) differential equation model of infections.

*Discrete Markov chain.*

This chain consists of 32 states which correspond to the states in the political territory division of the Mexican Republic, thus $E = \{e_i : e \in M, i \in \mathbb{N}\}$ where M stands for the Mexican territory, this set maps to $S = \{s_1, s_2, \ldots, s_{32}\}$ of discrete states $f: E \to S$ in a bijective fashion; let us call U this process, then $T \subset U$ where T stands for the set of temporal parameters, $T = \{t_j | j \in \mathbb{N}\}$ where the j-th stands for the j-th day or $[j] = [day]$, i.e., the corresponding unit of the temporal parameter is the day. Now let us establish the conditions of probability, $\forall s_i R s_k \exists p_{ik} \in \Omega$, for every existence of a relationship between the i-th and the k-th state there exists an associated probability *ik* for the transition from the i-th towards the k-th state, being this probability balanced between all the existent possible interactions between one i-th state and another any k-th state, for example, should the i-th state interact with n different states, then the associated probability for every $s_i R s_k$ is $p_{ik} = \frac{1}{n}$. The existence of an interaction between one state and another one is stated by means of the sharing of territorial boundaries. This process is featured by a stochastic matrix P,

$$P = \begin{pmatrix} p_{11} & p_{12} & p_{13} & \cdots & p_{1n} \\ p_{21} & p_{22} & p_{23} & \cdots & p_{2n} \\ p_{31} & p_{32} & p_{33} & \cdots & p_{3n} \\ \vdots & \vdots & \vdots & \ddots & \vdots \\ p_{n1} & p_{n2} & p_{n3} & \cdots & p_{nn} \end{pmatrix}$$

This matrix is shown in Fig. 1 where $\sum_{i=1}^{32} p_{1n} = 1$.

The initial vector of this process is one which displays the COVID19 features in Mexico by means of the report corresponding to 11/03/2020 issued by the Mexican Health Department, with the following cases: Mexico City, 5 cases; Coahuila, 1 case; Chiapas, 1 case; Sinaloa, 1 case (also a recovered case); Mexico State, 2 cases; Querétaro, 1 case. Thus, the total amount of cases can be described as the cardinality of the set of cases C, $|C| = 11$ and the probabilities of the initial vector $\pi^0$ for any given state are $\pi_i = \frac{\#cases}{|C|}$. The vectors then can be determined for any given discrete time $t_i \in T$,

$$\begin{aligned} \pi^0 &= \pi^0 \\ \pi^1 &= \pi^0 P^1 \\ \pi^2 &= \pi^0 P^2 \\ &\vdots \\ \pi^n &= \pi^0 P^n \end{aligned}$$

By computing the powers of the stochastic matrix we have computed the probability distributions when, $n = 2, 4, 8, 20, 40, 80, 160, 320$ reminding us that the units of this temporal parameter is $[t] = [day]$ thus we may compute the probability distributions of COVID19 for the following dates: 13/03/2020, 15/03/2020, 19/03/2020, 27/03/2020, 31/03/2020, 20/04/2020, 30/05/2020, 18/08/2020, 25/01/2020 whose graph is shown in Fig. 2.

When we compute the n-th power of the stochastic matrix we may realize as well of the stationary distribution of our set of variables or states, in this case, we may realize that there exists a tendency towards a stationary distribution, and the comparison between 25/01/2021 and 11/03/2020 is shown in Fig. 3, in which we may come to notice the decrease in inflexion points when $n = 320$ and the tendency towards a stabilisation, where the peaks of infection are reduced, and a homogeneous spread of the disease is witnessed. The probability distribution of every state when $n = 320$ is shown in Fig. 4; this last graph allows us to regard that the state with the highest probability or the highest prevalence at the given time is Campeche, followed by Quintana Roo; being the state with the lowest prevalence the Baja California Sur; but in no state do we observe $p = 0$. We have then successfully computed the probability distribution of every state in time, however we do not know the precise population in given time, the precise amount of total infected people, for which we will need the second model.

*Modified SIR model.*

This model allows us to know the amount of susceptible, infected and recovered people in our population set. The model will be derived from the following thesis,

$$S \xrightarrow{k_1} I \xrightarrow{k_2} R$$

Where S stands for susceptible population, I for infected population and R for recovered population; by means of chemical kinetics we may construct an analogy to these set of reactions,

$$-\frac{dS}{dt} = k_1 S$$

Now, let us consider $s_0$ as the initial susceptible population and $s_0 - x$ as the amount of susceptible people lost in the time of the disease,

$$S = s_0 - x$$

And we may say,

$$-\frac{dS}{dt} = -\frac{d(s_0 - x)}{dt}$$
$$= \frac{dx}{dt}$$

The corresponding differential equation is,

$$\frac{dx}{dt} = k_1(s_0 - x)$$
$$\frac{dx}{s_0 - x} = k_1 dt$$

Let us integrate this last expression,

$$\int \frac{1}{s_0 - x} dx = k_1 \int dt$$
$$\rightarrow \ln \frac{1}{s_0 - x} = k_1 t$$
$$\rightarrow \left[ \ln \frac{1}{s_0 - x} \right]_0^x = k_1 [t]_0^t$$

$$\rightarrow \ln \frac{s_0}{s_0 - x} = k_1 t$$

We obtain the following pieces of information,

$$x = s_0(1 - e^{-k_1 t})$$
$$S = s_0 - x$$
$$= s_0 e^{-k_1 t}$$

In order to compute the amount of infected people, we may define the equalities,

$$I = x - y$$
$$R = y$$

Then,

$$\frac{dy}{dt} = k_2(x - y)$$
$$= k_2(s_0(1 - e^{-k_1 t}) - y)$$

Which is a non-homogeneous linear differential equation,

$$\frac{dy}{dt} + k_2 y = k_2 s_0 (1 - e^{-k_1 t})$$

We may introduce the integrating factor,

$$\frac{d}{dt}[e^{k_2 t} y] = k_2 s_0 (1 - e^{-k_1 t}) e^{k_2 t}$$

We integrate,

$$e^{(k_2 t)} y = k_2 s_0 \left[ \int e^{k_2 t} y dt - \int e^{t(k_2 - k_1)} dt \right]$$
$$\rightarrow e^{k_2 t} y = k_2 s_0 \left[ \frac{1}{k_2} e^{k_2 t} - \frac{1}{k_2 - k_1} e^{t(k_2 - k_1)} \right]$$

By substituting limits,

$$[y]_0^y = \left[ s_0 - \frac{k_2 s_0}{k_2 - k_1} e^{-k_1 t} + e^{-k_2 t} \right]_0^t$$
$$\rightarrow y = e^{-k_2 t} - \frac{k_2 s_0}{k_2 - k_1} e^{-k_1 t} + \frac{k_2 s_0}{k_2 - k_1} - 1$$

Remembering,

$$I = x - y$$
$$= s_0 + e^{-k_1 t} \left( \frac{k_2 s_0}{k_2 - k_1} - s_0 \right) - e^{-k_2 t} - \frac{k_2 s_0}{k_2 - k_1} + 1$$

When we consider the initial susceptible population as,

$$s_0 = 146.54 x 10^6$$

And the constants to be,

$$k_1 = \frac{11}{s_0}$$
$$= 7.506 \times 10^{-8}$$
$$k_2 = \frac{1}{11}$$

We may estimate the proportions of the infected, susceptible and recovered populations in Mexico (Table 1).

### *Gather round!*

Let us now commute both models we have previously discussed in order to provide the cases' distribution in the states of our Markov chain (or the States of Mexico) (Fig. 5) where we may notice the slow infective rate of COVID19, and a tendency to spread in every state in Mexico where by n=320, there will be $c \geq 1$ cases per state, should there be no preventive measures taken.

Addenda.

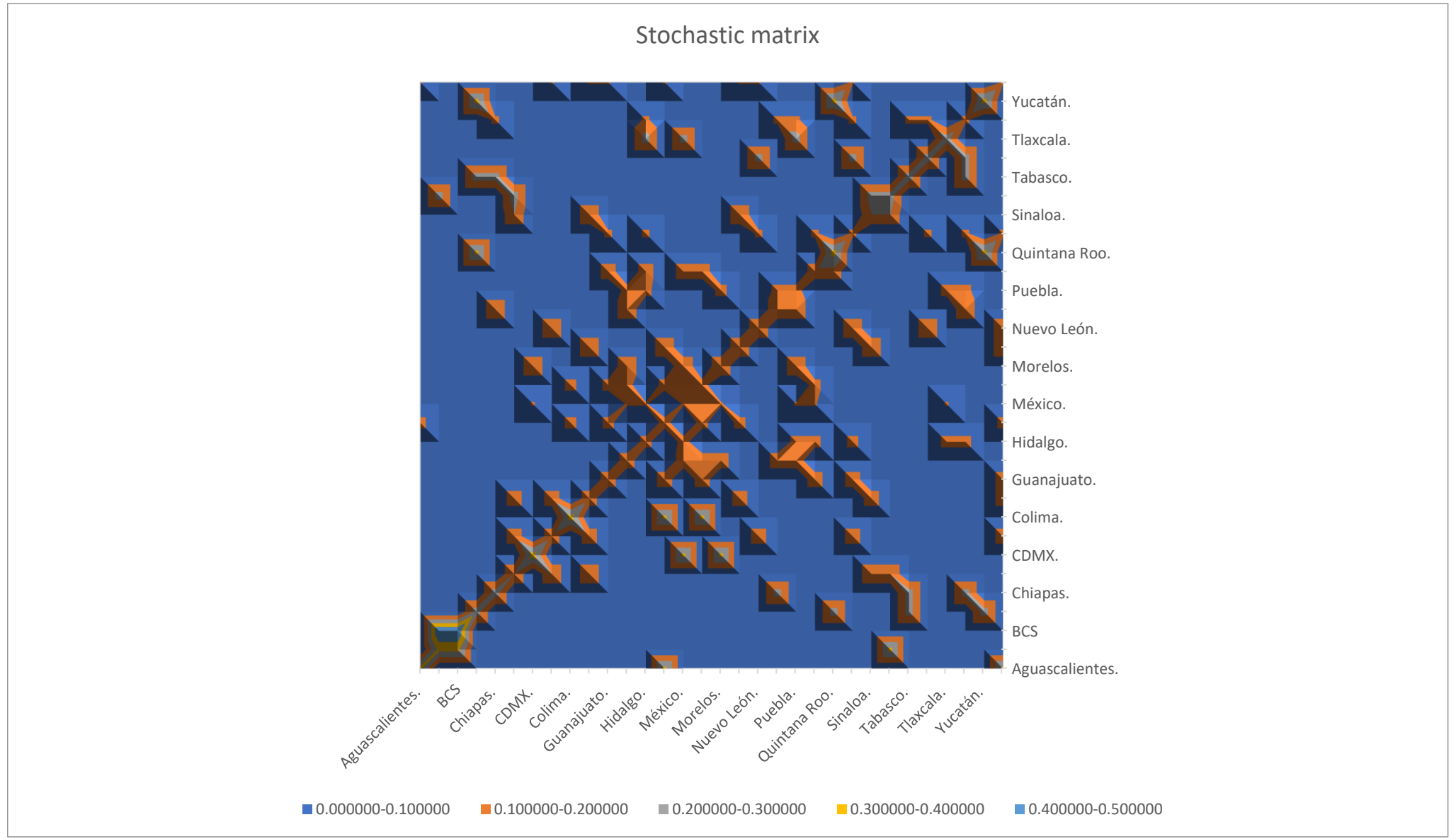

Fig. 1. Graph showing the stochastic matrix corresponding to the transition probabilities between states.

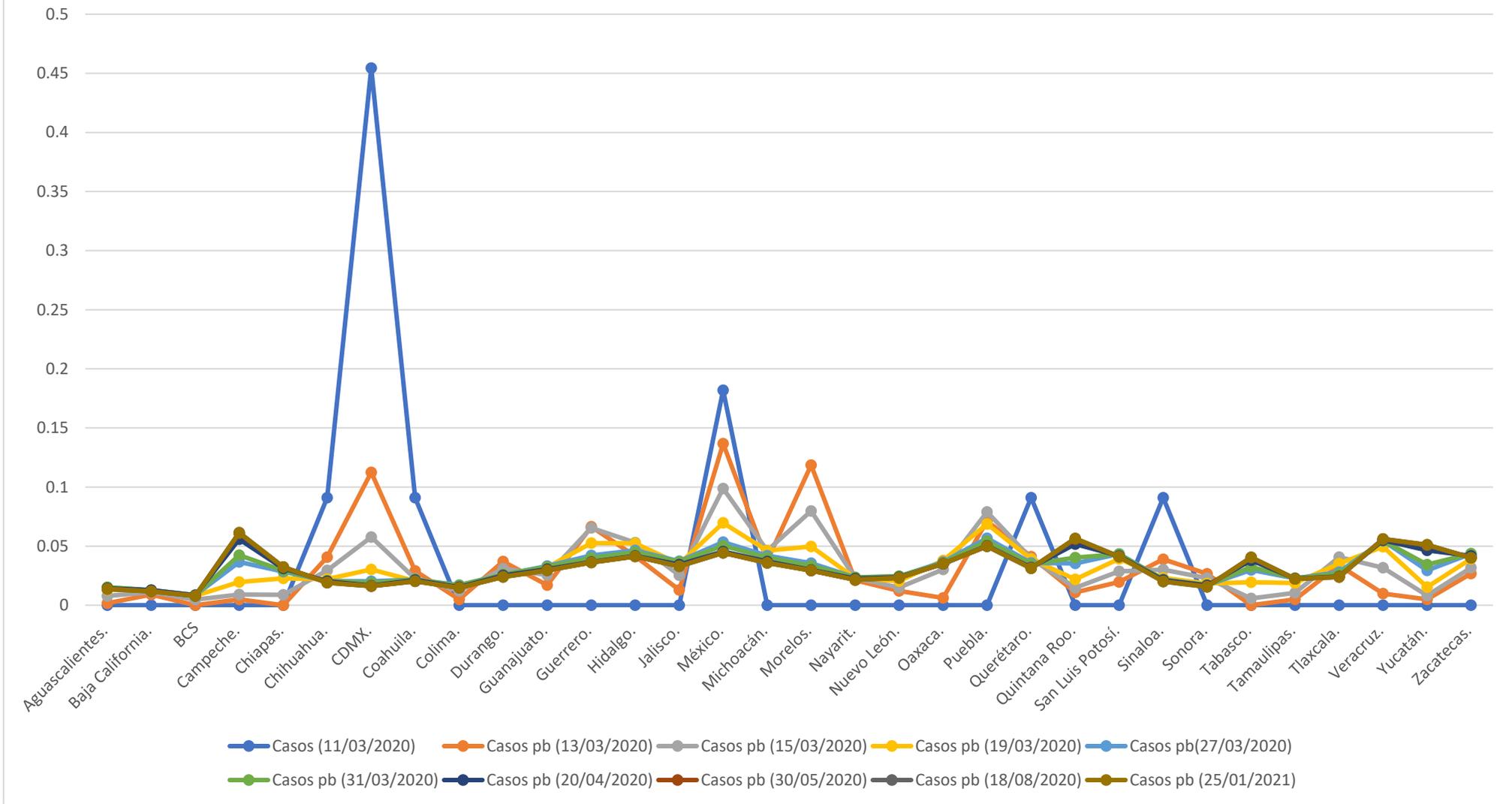

Fig. 2. Graph showing the probability distribution of COVID19 in a temporal perspective until 25/01/2021.

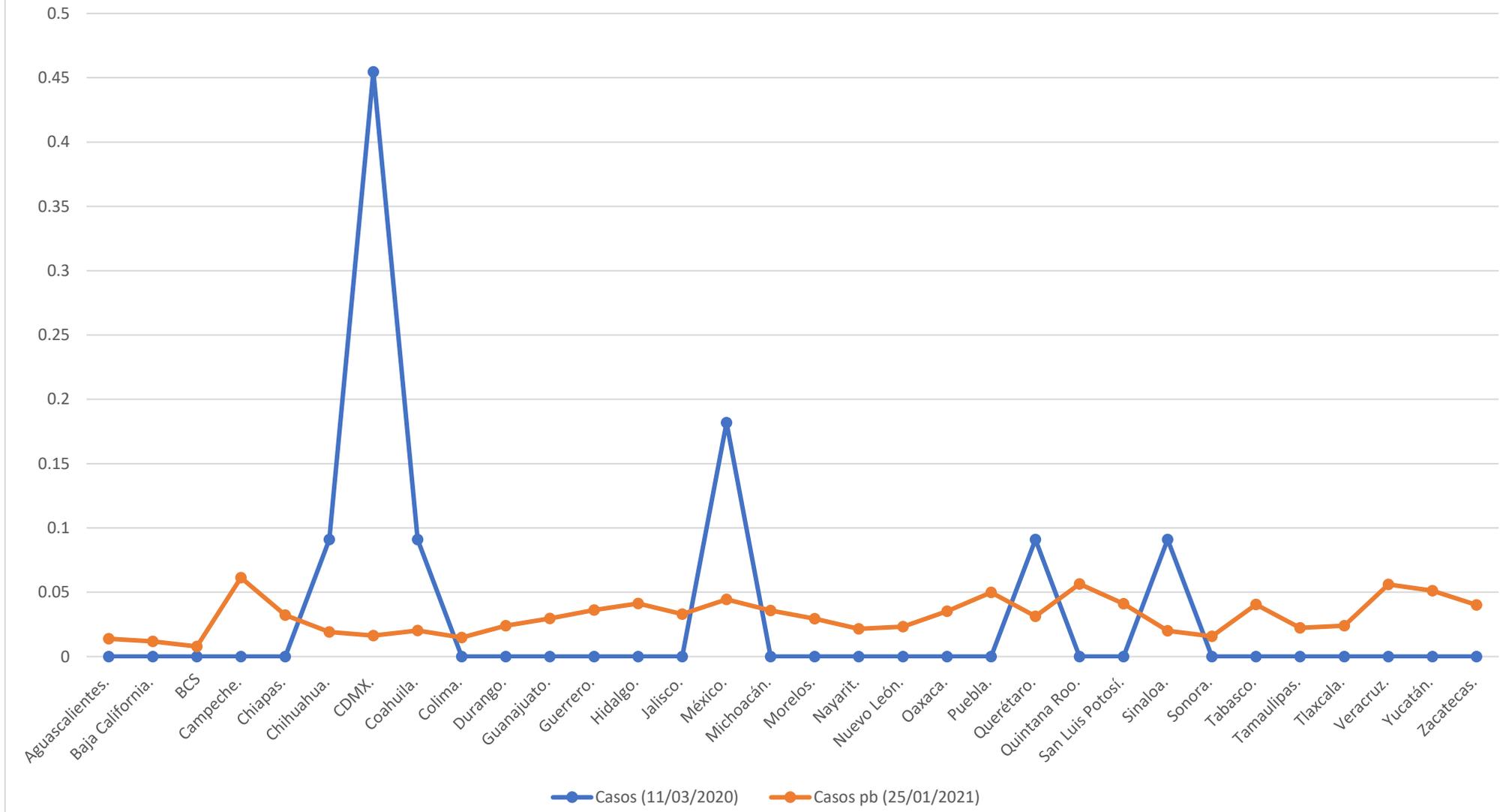

Fig. 3. Graph comparing the probability distribution of COVID19 in Mexican population between 11/03/2020 and the last measurement made when $n = 320$ or 21/01/2021.

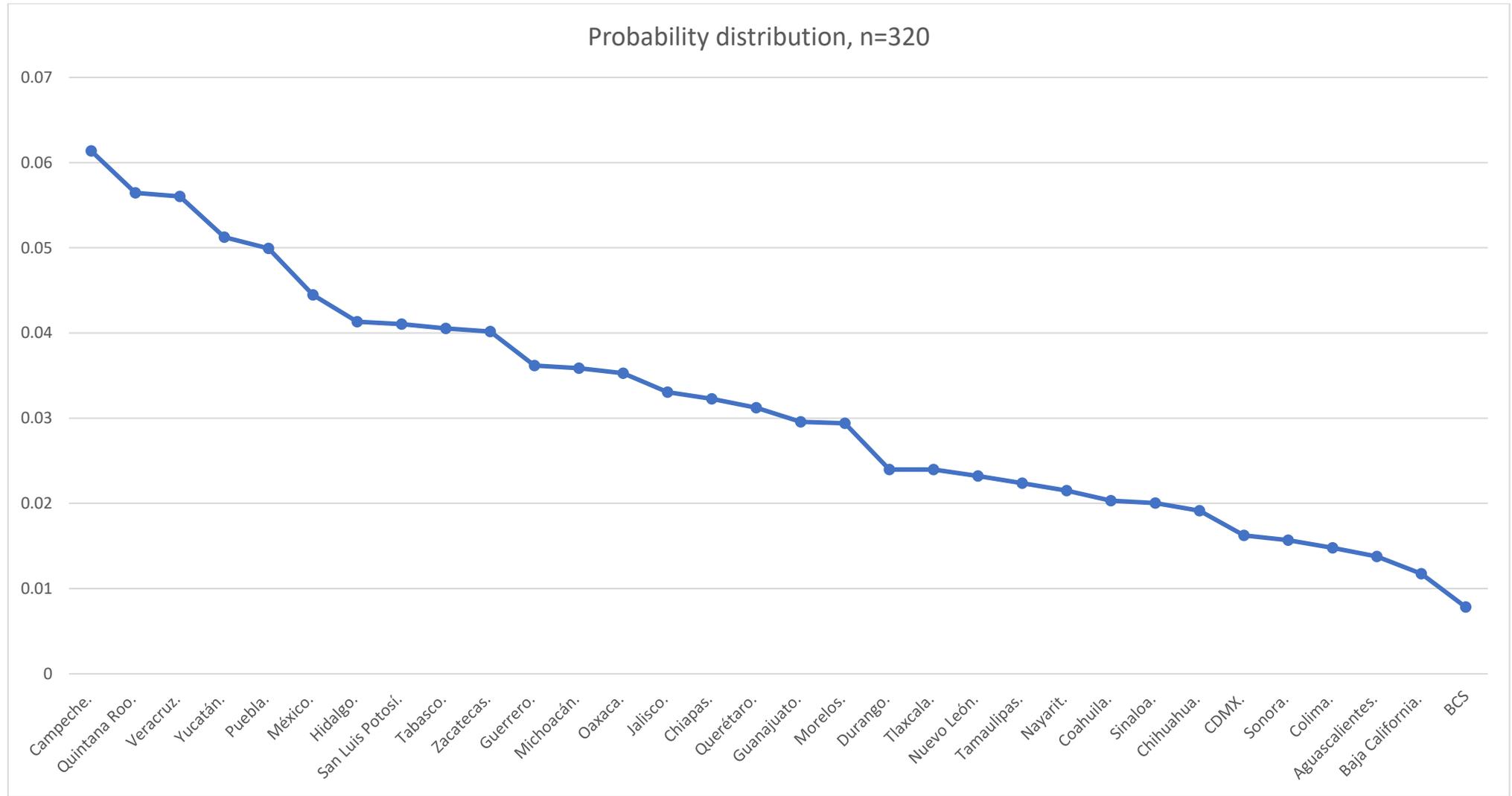

Fig. 4. Probability distribution of every state when n=320.

|  | I | S | R |
|---|---|---|---|
| Cases (11/03/2020) | 11 | 146540000 | 1 |
| Cases pb (13/03/2020) | 21.9448301 | 146539978 | 0.05375291 |
| Cases pb (15/03/2020) | 43.8620184 | 146539956 | 0.1351439 |
| Cases pb (19/03/2020) | 87.6310877 | 146539912 | 0.36322509 |
| Cases pb (27/03/2020) | 109.489992 | 146539890 | 0.50289033 |
| Cases pb (31/03/2020) | 218.623361 | 146539780 | 1.36232062 |
| Cases pb (20/04/2020) | 436.544687 | 146539560 | 3.42634798 |
| Cases pb (30/05/2020) | 872.141055 | 146539120 | 7.79969422 |
| Cases pb (18/08/2020) | 1743.27722 | 146538240 | 16.5990005 |
| Cases pb (25/01/2021) | 3485.5343 | 146536480 | 34.197 |

Table 1. Number of estimated cases: infected (I), susceptible (S), recovered (R) people throughout time (Cases pb=probable cases).

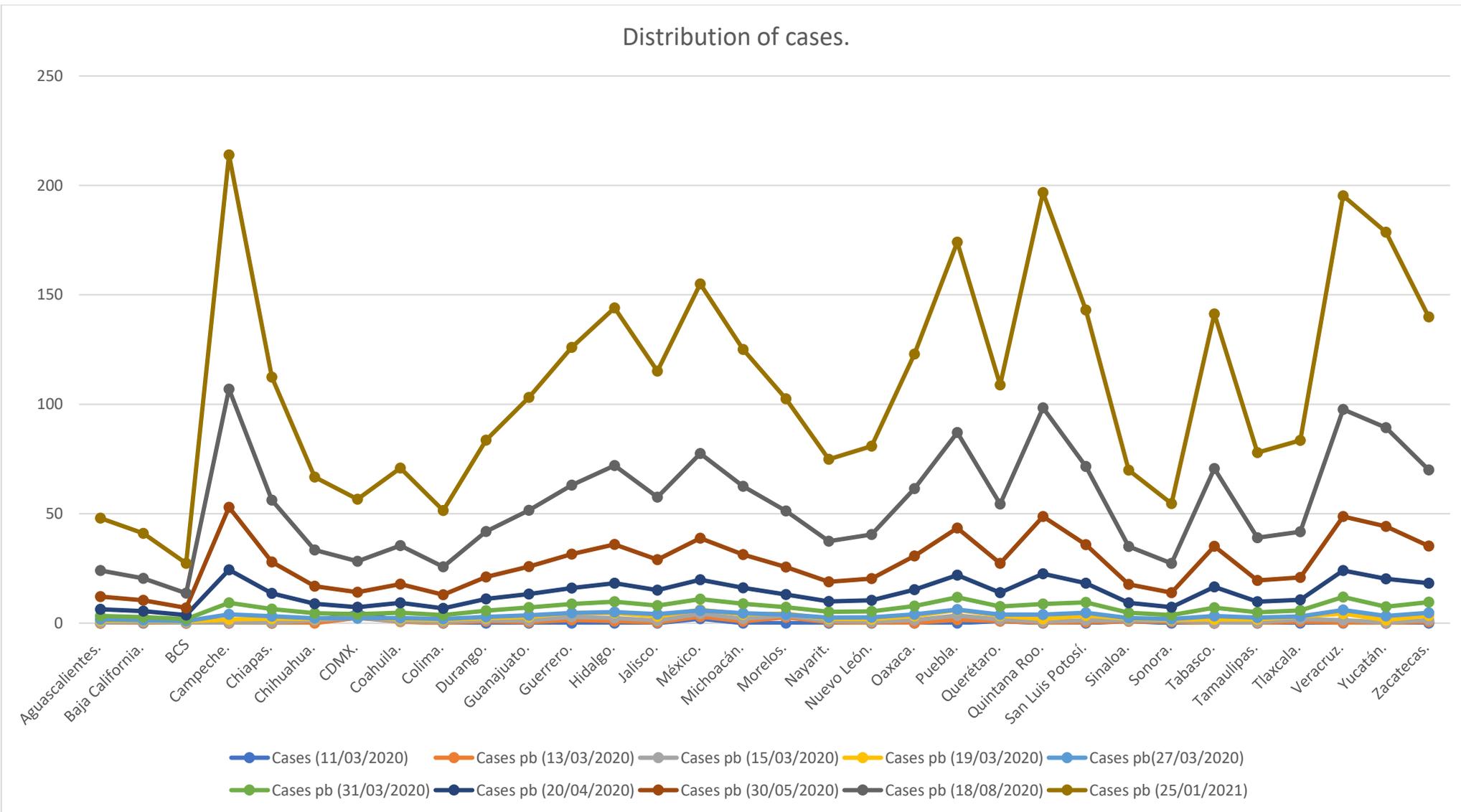

Fig. 5. Graph displaying the distribution of estimated cases computed through the SIR model and the probabilities' distribution by means of the Markov chain.